\documentclass[sigconf,natbib=true]{acmart}
\AtBeginDocument{%
  \providecommand\BibTeX{{%
    \normalfont B\kern-0.5em{\scshape i\kern-0.25em b}\kern-0.8em\TeX}}}

\copyrightyear{2023}
\acmYear{2023}
\setcopyright{rightsretained}
\acmConference[SIGIR-AP '23]{Annual International ACM SIGIR Conference on Research and Development in Information Retrieval in the Asia Pacific Region}{November 26--28, 2023}{Beijing, China}
\acmBooktitle{Annual International ACM SIGIR Conference on Research and Development in Information Retrieval in the Asia Pacific Region (SIGIR-AP '23), November 26--28, 2023, Beijing, China}
\acmDOI{10.1145/3624918.3625335}
\acmISBN{979-8-4007-0408-6/23/11}

\usepackage{xspace}
\usepackage{multirow}
\usepackage{graphics}
\usepackage{amsfonts}
\usepackage{subcaption}
\usepackage{enumitem}
\usepackage{float}
\usepackage{listings}
\usepackage{xcolor}

\begin{document}

\newcommand{\ourdataset}{\texttt{SE-PEF}\xspace} 

\title{\ourdataset: a Resource for Personalized Expert Finding}

\author{Pranav Kasela}
\affiliation{
  \institution{University of Milano-Bicocca}
  \city{Milan}
  \country{Italy} \and
  \institution{ISTI-CNR, Pisa, Italy}
  \city{Pisa}
  \country{Italy}
}
\email{pranav.kasela@unimib.it}

\author{Gabriella Pasi}
\affiliation{
  \institution{University of Milano-Bicocca}
  \city{Milan}
  \country{Italy}
}
\email{gabriella.pasi@unimib.it}

\author{Raffaele Perego}
\affiliation{%
  \institution{ISTI-CNR, Pisa, Italy}
   \city{Pisa}
   \country{Italy}
   }
 \email{raffaele.perego@isti.cnr.it}

\renewcommand{\shortauthors}{Kasela, et al.}

\begin{abstract}
The problem of personalization in Information Retrieval has been under study for a long time. A well-known issue related to this task is the lack of publicly available datasets to support a comparative evaluation of personalized search systems. To contribute in this respect, this paper introduces \ourdataset (StackExchange - Personalized Expert Finding), a resource useful for designing and evaluating personalized models related to the Expert Finding (EF) task.
The contributed dataset  includes more than  250k queries and 565k answers from 3 306 experts, which are annotated with a rich set of features modeling the social interactions among the users of a popular cQA platform. 
The results of the preliminary experiments conducted show the appropriateness of \ourdataset to evaluate and to train effective EF models. 

\end{abstract}







\maketitle

\section{Introduction}

Expert finding (EF) is a well-studied problem in community question answering (cQA).
The aim of EF in a cQA scenario, is to identify users, namely the experts, that might be able to answer correctly a given question on a specific topic.
This task is  important for many applications, e.g., crowd-sourcing, and for cQA platforms that wish to increase  user engagement by precisely identifying the experts to whom to  propose the questions about a given topic. 

Personalization is gaining traction in many IR \cite{pasi2013pers, bassani2022multi, bassani2023denoising, ma2020pstie} and NLP \cite{braga2023PersonalizationIB} tasks, but it is not largely adopted in EF due to the lack of publicly-available, large-scale datasets containing user-related information.
In this research paper, building upon our previous work~\cite{se-pqa}, in which the authors presented a dataset for personalized community question answering, we introduce \ourdataset (StackExchange - Personalized Expert Finding), a large dataset rich in user-level features that can be leveraged for training, evaluating and comparing both personalized and non-personalized models for the \textit{Expert Finding} task.
\ourdataset comprises around 250k questions and 560k associated answers provided by 3,306 experts, and it inherits a rich set of features modeling the social interactions within the user community.  
To train personalized models, we keep the user-related data as they are provided in the original dataset: users' past questions and answers, their own social autobiography, their reputation score, and the number of profile views that they have received.

In the case of EF, personalization can improve the perceived service quality in different ways. 
For example, when the requesting user is interested in multiple topics, identifying an expert by considering also the requesting user's interests can improve the trust in the answer received. 
A similar effect can be obtained by preferring experts that are closer to the requesting users  based on past interactions or follower/followee dynamics.
In summary, the contribution of this paper is the following:
\begin{itemize}[leftmargin=*]
    \item We provide and make available the \ourdataset dataset, as a public resource consisting of a comprehensive corpus including around 255k questions and 560k answers provided by $3 306$ expert users. The richness and variety of features provided with the dataset enable its use for the design and evaluation of personalized EF methods.
    
    \item We report a preliminary comparison of the performance of different EF methods applied to the questions, answers, and users in \ourdataset. The results confirm that models based on deep learning outperform in  effectiveness  traditional retrieval models and that by exploiting personalization features we can obtain a significant performance boost.
\end{itemize}
The rest of the paper is organized as follows. 
Section \ref{sec:dataset} introduces the \ourdataset dataset and reports some statistics about its content. Moreover, the section details the EF  task addressed in this paper by using   \ourdataset. 
Section \ref{sec:available} compares \ourdataset with respect to other publicly-available resources in the  field. 
Section \ref{sec:experiments} presents a preliminary comparison of traditional and personalized models for EF applied to \ourdataset. 
In Section \ref{sec:utility} we discuss the utility and the practical implications of the new resource.  
Finally, Section \ref{sec:conclusion} concludes the work and draws future lines of investigation.

\section{The \ourdataset Dataset}
\label{sec:dataset}

The dataset proposed in \cite{se-pqa} is based on StackExchange\footnote{\url{https://stackexchange.com}} and available under a CC BY-SA 4.0 license.
It comprises questions and answers from 50 different stackexchange communities, written between  \textit{2008-09-10} and \textit{2022-09-25}. There are around 1.1 million questions and 2.1 million answers. The training, validation and test splits are based on a temporal condition and are already provided on zenodo\cite{z_kasela2023sepqa}. 

In \cite{se-pqa} the authors show that personalization is more useful if multiple communities are used together in this dataset rather than using a single community to create the dataset.
Meanwhile, previous works that use StackExchange for EF tasks focus only on a single community or a portion of a community, thus neglecting  the domain diversity characterizing the questions and the various experts~\cite{stackoverflow_expert, stackexchange_expert1, stackexchange_expert2}. 

\begin{figure}
    \centering
    \includegraphics[width=\linewidth]{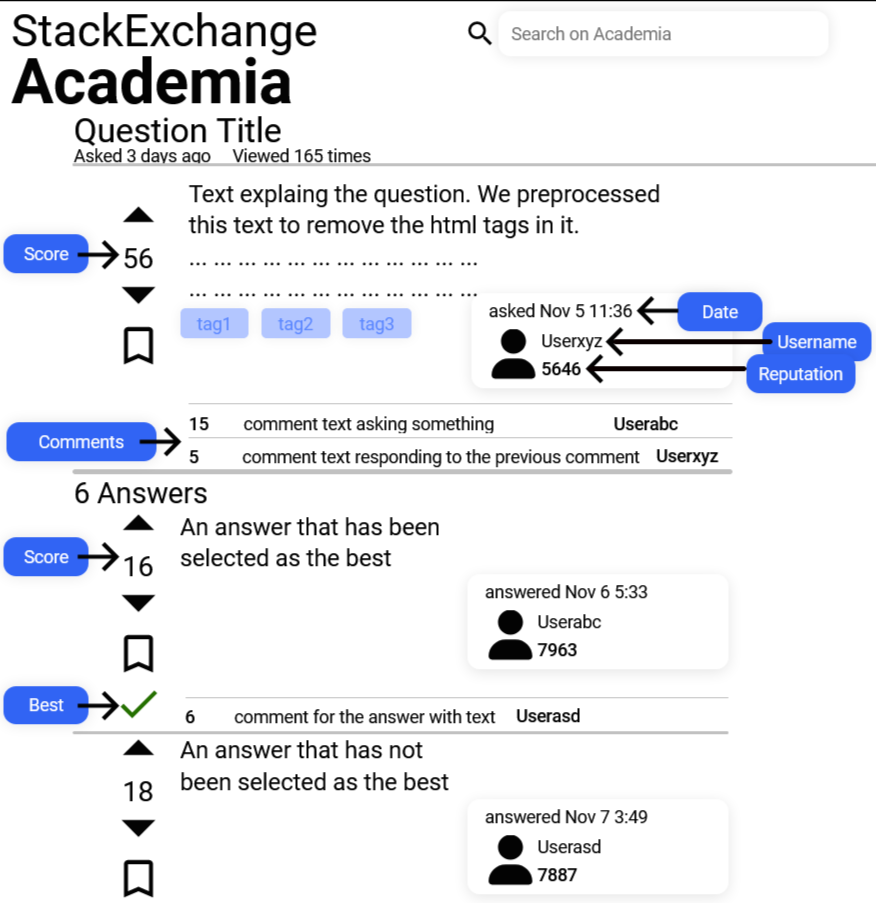}
    \caption{Illustration of StackExchange data.}
    \Description{UI example of the StackExchange website. The figure highlights the position of some of the most important features available, for example, post score, comment location, date of publication, username.}
    \label{fig:stack_ui}
\end{figure}

\subsection{Accessing the \ourdataset dataset}

\ourdataset dataset is made publicly available on zenodo\footnote{\href{https://doi.org/10.5281/zenodo.8332747}{https://doi.org/10.5281/zenodo.8332747}\cite{z_kasela2023sepef}} according to the conditions detailed in the included CC BY-SA 4.0. license agreement and the code used for data creation, training, hyper-parameter optimization, and testing are available on github\footnote{\href{https://github.com/pkasela/SE-PEF}{https://github.com/pkasela/SE-PEF}}.

\subsection{\ourdataset Definition}
\label{subsec:taskdefinition}

In the following, we introduce the specific instance of EF task in which we are interested and illustrate how to address it by using the resources in \ourdataset.

Our EF task shares the same goal as the question-answering task: satisfy users' needs in a cQA forum in the most effective way. 
In a cQA forum, a user may ask a question that does not have any related answers in the answer collection.
Since not receiving any answer can create a sense of frustration in a user posting a question, it is important for the community and the platform to identify and, eventually, notify domain experts who may be able to answer the question correctly. 
Finding good matches between unanswered questions and  expert users can improve remarkably the engagement with the community. In fact, on the one hand, users posting a question can receive correct answers from the alerted experts in a short time; on the other hand, expert users  can dedicate their time to answering questions specifically related to their expertise  rather than searching for  questions that they can respond.

Formally, let $\mathcal{E}$ be a set of expert users $\{e_1, \ldots, e_k\}$. Given a question $\mathbf{q}$ asked by user $\mathbf{u}$, the EF task consists in retrieving from $\mathcal{E}$ a list of  $k$ experts $\{e_{q,1}, \ldots, e_{q,k}\}$ ordered by their likelihood of answering correctly to question $\textbf{q}$.

StackExchange data 
has been used in several EF papers, e.g., in~\cite{stackexchange_expert1, stackexchange_expert2, stackoverflow_expert}. These works however mostly focus on solving the expert finding task for a single community. 
\ourdataset incorporates instead information from multiple communities to provide a dataset that  can be used also to investigate models for generalist cQA forums that may not have separate channels for the discussed topics.

To create the dataset, we define as \textit{best answer} for a given question the answer selected as the best one by the user who asked the question, if available; otherwise, we assume the best answer to the one with the highest score, if it has received a score greater than a fixed threshold $\gamma_s$ \footnote{The $\gamma$ thresholds  used for \ourdataset are reported at the end of Section \ref{sec:available}.}. 
We note that this assumption, for the best answer being the most voted answer if no answer has been flagged as best by the user asking the question, is used only for the expert detection procedure, which will be explained subsequently and not as relevance judgement for the test data.
In the test set we only consider the best answer, the answer explicitly labeled as such by the user asking the question.  
Exploiting high-scored answers as the best answers allows us to increase the number of questions successfully answered. Indeed this choice is justified by the observation that  87.6\% of the answers, which are selected as best ones by the user asking the question, are also the most up-voted ones. On the other hand, we have observed that many users, once they satisfy the information need with a good answer, do not bother to mark the answer as the best.\\

At this point, to identify the set of experts  $\mathcal{E}$, we follow the procedure indicated by Dargahi et al.~\cite{stackoverflow_expert} for their StackOverflow dataset:
\begin{itemize}
    \item For each community $C$, let $\mathcal{U}$ be the set of  users, and $\mathcal{B}$  the set of  best answers computed as explained above in the community $C$. For each user $\mathbf{u} \in \mathcal{U}$,
    let $\mathcal{A}_{u, C} = \{ a_{u,1}, a_{u,2}, \hdots, a_{u,n} \}$ be the set of answers given by $\mathbf{u}$ in $C$;
    \item Remove all users who do not have at least $\gamma_a$ answers selected as best answers, i.e. define:
    \begin{displaymath}
        \mathcal{E}' = \mathcal{U}\setminus \{u \in \mathcal{U}, \text{ s.t. } |\mathcal{A}_{u, C} \cap \mathcal{B}| < \gamma_a \}
    \end{displaymath}
    \item Compute the acceptance rate for the  users in $\mathcal{E'}$ given by the ratio between the number of accepted answers and the number of total answers of the user in that community. 
    For each user $\mathbf{e'}$ we define $ar_{u,C}$:
    \begin{displaymath}
        ar_{e', C} = \frac{|\mathcal{A}_{e', C} \cap \mathcal{B}|}{|\mathcal{A}_{e', C}|}
    \end{displaymath}
    \item Compute the average acceptance rate $\bar{ar}_C$ for the users in a community and select as experts only those users who have an acceptance rate  above the community average one:
    \begin{displaymath}
        \mathcal{E}_C = \{e \in \mathcal{E}' \text{ s.t. } ar_{e', C} \geq \bar{ar}_C\}
    \end{displaymath}
\end{itemize}
The final set of experts $ \mathcal{E}$ is defined as the union of the sets of experts found for each community. 
The above  process ensures that the selected experts have a high level of engagement and write high-quality answers having a high acceptance rate.

\begin{figure*}[htbp]
    \centering
    \lstset{
        string=[s]{"}{"},
        stringstyle=\color{blue},
        comment=[l]{:},
        commentstyle=\color{black},
    }
    \begin{lstlisting}
    {
      "id": "academia_49906",
      "text": "Including teaching statement in RA position application package [...]",
      "timestamp": 1438693784,
      "user_id": 3422261,
      "user_questions": ["academia_28238", "academia_28240", ...],
      "user_answers": ["genealogy_4058", "academia_37538", ...],
      "tags": ["application"], 
      "expert_ids": [339125], 
      "expert_questions": [[]],
      "expert_answers": [["expatriates_2520", "academia_18991", ...]]
    }
    \end{lstlisting}
    \caption{Example of a line of the jsonl file provided.}
    \label{fig:data_example}
\end{figure*}

In Figure \ref{fig:data_example} we show the basic structure of the JSON file provided for training, validation, and test. 
The user\_questions, user\_answers contain the identifiers (ids) of the questions and the answers, written before the current question timestamp, of the user asking the question. 
The expert\_questions, and expert\_answers contain the ids of the questions and the answers of the expert that has given the best answer.
The data is provided also with a collection of questions and a collection of answers; they are two very simple JSON files, where the keys correspond to the ids of the questions and answers respectively. The values of the keys are constituted by the texts of the questions and answers respectively.  
The data is provided also with multiple data-frames, curated from the original data found from archive.org, which can be used to add more features.
These features are described on the Stack Exchange website.\footnote{\href{https://meta.stackexchange.com/questions/2677/database-schema-documentation-for-the-public-data-dump-and-sede}{https://meta.stackexchange.com/questions/2677/database-schema-documentation-for-the-public-data-dump-and-sede}}
\section{Comparison with available datasets}
\label{sec:available}

\begin{table}
    \centering
    \caption{Comparison between \ourdataset and other cQA  datasets for EF. When a specific definition of an expert is provided we distinguish  normal users from experts.}
    \label{tab:expert_compare}
    \resizebox{\linewidth}{!}{
    \begin{tabular}{lcccc}
         \toprule
         Dataset & Questions & Answers & Users & Experts\\
         \midrule
         StackOverflow & 123 933 & N/A & 22 027 & 1845\\
         Quora & 444 138 & 887 771 & 95 915 & N/A\\
         Wondir D5 
         &  \multicolumn{2}{c}{752 391 QA pairs} & N/A & 17 525\\
         Wondir D20 &  \multicolumn{2}{c}{639 233 QA pairs} & N/A & 5 025\\
         Yahoo U10 
         & 32 009 & 97 911 & 2 515 & N/A\\
         Yahoo U15 & 28 404 & 89 144 & 1 339 & N/A\\
         Yahoo U20 & 25 690 & 80 677 & 870   & N/A\\
         StackExchange$_{\text{Gis}}$ & 50 718 & 70 034 & N/A & 3 168\\
         StackExchange$_{\text{English}}$ & 46 692 & 104 453 & N/A & 4 781\\
         StackExchange$_{\text{CodeReview}}$ & 36 947 & 57 622 & N/A & 2 242\\
         
         \textbf{\ourdataset} & \textit{255 352} & \textit{564 690} & \textit{81 252} & \textit{3 306}\\ 
         \bottomrule
    \end{tabular}
    }
    
\end{table}

Concerning the EF task, there are plenty of datasets available~\cite{expert_finding_survey}, and some of them are based on data from cQA websites. For example, StackExchange is used to create a pre-trained BERT model for the EF task in \cite{stackexchange_expert2}. However,  the  work focuses only on designing an EF pre-training framework based on a specific augmented masked language model able to learn the question-expert matching task. 
Other EF datasets derived from cQA forums come from: StackOverflow~\cite{stackoverflow_expert,stackoverflow_expert2}, Yahoo Answers!\cite{yahooexpert,yahooexpert2}, Wondir~\cite{wondirexpert} and Quora~\cite{quoraexpert}.
Recently, a  domain-specific expert finding task was tackled using Avvo~\cite{askari2022expert}, a legal cQA website, but in this case, personalization is not possible due to the fact that users are anonymous.
In Table \ref{tab:expert_compare} we report the basic dataset statistics of some of the commonly used datasets in EF for comparison.

A common issue with the existing datasets is that the experts are, in many cases, not well-defined, and determining what makes a user an expert is not trivial.
Furthermore, most works among those previously cited  either  rely on a private dataset, or refer to a specific domain and make very strong assumptions simplifying the task addressed.  
Conversely, \ourdataset will be made publicly available, it has a well-defined definition of an expert, which is inspired by reasonable hypothesis common to other works~\cite{stackexchange_expert1,stackexchange_expert2,stackexpert20}.
Furthermore, it provides  a rich set of social features usable for personalization and combines data from multiple communities, which, as we have already stated, increases  dataset diversity and opens the possibility of exploiting cross-domain user information for EF.  

To build the \ourdataset for EF we followed the procedure detailed in Section \ref{subsec:taskdefinition}, by setting $\gamma_s = 5$ and $\gamma_a = 10$. Finally, we also remove from the training dataset the questions answered by  experts who previously posted less than 5 answers to avoid the cold start problem for expert modeling.
Using this procedure, we obtain \ourdataset, from starting from the dataset presented in \cite{se-pqa}, including 81,252 users, 3,306 experts, 252,501 queries (218,647 for training, 16,710 for validation, and 19,995 for testing), and 564,690 answers.

\section{Preliminary experiments with \ourdataset}
\label{sec:experiments}

This section provides a concise overview of the experimental setup and introduces the methods employed to showcase the capabilities of \ourdataset 
in the EF task, defined and discussed in  Section \ref{subsec:taskdefinition}. 
Finally, we report and discuss the results of the conducted experiments.

\subsection{Experimental settings}

For our EF task, we use a retrieval-based approach~\cite{expert_voting}, and simply cast the EF task to a cQA task where we use the similarity scores of the retrieved documents as experts' scores. We explain this in detail in the following paragraphs. 

We adopt a two-stage ranking architecture that prioritizes efficiency and recall in the first stage. The primary objective of this first stage is to select for each query a set of candidate documents that are eventually re-ranked in a second stage by a precision-oriented ranker.
The first stage is based on Elastic Search\footnote{https://www.elastic.co/}, and uses  BM25 as a fast ranker. We use the same BM25 hyperparameters as indicated in \cite{se-pqa}: 1 and 1.75 for b and k1, respectively.
In the second, precision-oriented stage,  to re-rank
the retrieved documents we utilize a linear combination of  the set of available scores that includes the BM25 score, the similarity score computed by a neural re-ranker, and, when used, the score computed by a personalization model exploiting the user history.  In all the experiments the second stage re-ranks the top-100 results retrieved with BM25.

\paragraph{Non-personalized models}
As neural re-ranker in the second stage we use the following two  models used also  in \cite{se-pqa}:
\begin{itemize}
    \item  DistilBERT. This model is obtained by fine-tuning the pre-trained \textit{distilbert-base-uncased} model\footnote{https://huggingface.co/distilbert-base-uncased} for the task of answer retrieval tackled in \cite{se-pqa}.
    We use the same training data and experimental settings used in \cite{se-pqa}.
  
    \item MiniLM, based on MiniLM-L6-H384-uncased\footnote{https://huggingface.co/sentence-transformers/all-MiniLM-L6-v2}. This model is used as it is, without any fine-tuning. 
\end{itemize}

\paragraph{Personalized model for EF} 

For building the EF personalized model we exploit the folksonomy arising from tags, very similar to the one employed in \cite{se-pqa}. 
This model, which we also call TAG from hereon, aims at capturing the similarities among the topics addressed by the asker in their current and previous questions, and the ones in which a considered expert answered in the past.
Given a question $\textbf{q}$, asked by user $\textbf{u}$ at time $\textbf{t}$, let $T_{u,t}$ be the set of tags assigned by $\textbf{u}$ to all theirs questions posted before $\textbf{t}$ (including $\textbf{q}$). 
$T_{u,t}$ thus represents the interests of $\textbf{u}$ as expressed in their previous interactions. 
The authors of the answers to query $\textbf{q}$ do not have the possibility of tagging explicitly their answers, so for each answer, we consider  the tags associated with the answered questions. 

The way we represent the expert user is slightly different: the expertise, in this case, is based on  a pre-computed, static representation $T_e$ of each expert $e$ in \ourdataset. This representation considers the tags $T'_e$ of all the questions answered by $e$  included only in the training set. To build $T_e$ from $T'_e$ we perform an additional step consisting in discarding the tags with a frequency lower than the median frequency of tags in $T'_{e}$.  This tag pruning step reduces the noise coming from the possible presence of non-relevant tags that might have appeared as additional tags in a few  questions answered but the expert might not be an expert on those topics. As for the previous task, 
the EF TAG score $s_{e,q}$ for expert $e$ is finally computed as
\begin{displaymath}
    s_{e,q} = \frac{|T_e \cap T_{u,q}|}{|T_{u,q}| + 1}
\end{displaymath}    

\paragraph{Score computation and combination}
Given the list of answers $A$ retrieved and re-ranked with the  above models, we observe that some users  could have authored multiple answers included in $A$.
This is potentially an important feature for characterizing their expertise.
Therefore, to obtain the expert-level score, we sum up the scores assigned to all the answers in $A$ coming from the same expert. Moreover, since the TAG model returns a score for all experts in the dataset, even those not having an answer in $A$, we assume that these experts receive a score contribution equal to $0$ from BM275 and the non-personalized models. 
Finally, the scores from BM25, personalized and non-personalized models are combined   by computing  the weighted sum of the normalized scores from the  models, using  weights $\lambda_{BM25}, \lambda_{DistilBERT/MiniLM},$ and $\lambda_{TAG}$,  with $\sum_i \lambda_i = 1$.
The  $\lambda$ values are optimized on the validation set by performing a grid search in the interval $[0, 1]$ with step $0.1$.

\paragraph{Evaluation Metrics}
For the task of expert finding we utilize the following evaluation metrics: Precision at 1 (P@1), Recall at 3 (R@3), Recall at 5 (R@5), and Mean Reciprocal Rank at 5 (MRR@5) as our evaluation metrics.
The cutoffs are set low as we prioritize identifying experts at the top of the ranked lists.
All the metrics are computed by using the \textit{ranx} library~\cite{bassani2022ranx, ranx_fusion}. 

\subsection{Experimental Results}
\label{subsec:exResults}

\begin{table}
    \centering
    \caption{Results for the \ourdataset EF task.}
    \label{tab:expert_result}
    \resizebox{\linewidth}{!}{
    \begin{tabular}{lllllc}
        \toprule
        Model & P@1 & R@3 & R@5 & MRR@5 & $\lambda$\\
        \midrule
        BM25 & 0.134 & 0.255 & 0.314 & 0.200 & - \\
        BM25 + TAG & 0.150* & 0.286* & 0.361* & 0.226* & (.4,.6) \\
        \midrule
        MiniLM$_{\text{SBERT}}$ & 0.126 & 0.238 & 0.296 & 0.188 & - \\
        MiniLM$_{\text{SBERT}}$ + TAG & 0.143* & 0.276* & 0.348* & 0.217* & (.4, .6)\\
        \midrule
        DistilBERT & 0.147 & 0.274 & 0.334 & 0.216 & - \\
        DistilBERT + TAG & \textbf{0.163}* & \textbf{0.304}* & \textbf{0.375}* & \textbf{0.240}* & (.5, .5)\\
        \bottomrule
    \end{tabular}
    }
\end{table}

The results are reported instead in Table \ref{tab:expert_result}. 
The symbol * indicates a statistically significant improvement over the respective non-personalized method not using any contribution from the TAG model. Statistical significance is assessed with a Bonferroni-corrected two-sided paired student's t-test with 99\% confidence. 
The column labeled $\lambda$ reports the optimized weights, found using the validation set, used for combining the scores computed by BM25, DistilBERT / MiniLM, and TAG models.
In the cases in which the optimal weight for the BM25 score is equal to $0$ -- i.e., BM25 does not contribute to re-ranking -- we omit  BM25 from the name of the model and $\lambda_1 = 0$ from the weights column.\\

Differently from the cQA task tackled by the authors of \cite{se-pqa}, we observe that on EF the performance gap of DistilBERT vs. MiniLM$_{\text{SBERT}}$ is sensibly reduced. 
The best-performing model among the ones tested is in fact DistilBERT + TAG which significantly outperforms both DistilBERT and MiniLM$_{\text{SBERT}}$. 
Analogously to the cQA task, personalization is very effective for EF. The contribution of the  TAG model allows for significantly improving all the non-personalized methods, with a performance boost exceeding three points in  MRR@5 for the DistilBERT model. By looking at the optimized $\lambda$ weights reported in all three tables, we see that the TAG model contribution is much higher for the EF task ($\lambda_{TAG} \geq .5$) than for the one obtained by the authors of \cite{se-pqa} ($\lambda_{TAG} \leq .3$).

\section{Utility and predicted impact}
\label{sec:utility} 
The \ourdataset resource we make available to the research community  is a step ahead toward a fair and robust evaluation of personalization approaches in Expert Finding.
The features inherited from \cite{se-pqa} include explicit signals to create relevance judgments and a large amount of historical user-level information  to design  and test classical and novel personalization methods.\\
We expect the \ourdataset dataset being useful for many researchers and practitioners working in personalized IR and the application of machine/deep learning techniques for personalization. 
In recent years, significant efforts have been dedicated to the study of personalization techniques. 
However, there is still a lack of a comprehensive dataset for evaluating and comparing different approaches, which makes the comparison between different methods less reliable or, worse, not possible at all. 

For this reason, we expect that the proposed dataset will impact the  research community working on personalized EF as it provides a common ground of evaluation built on questions, answers, and experts from real users socially interacting via a community-oriented  web platform. 

In this proposal, the expert can have different domain backgrounds and share interests and knowledge in various communities.
We also expect that training training on such rich and diverse data, like \ourdataset, should produce a more robust and generalizable model.

\section{Conclusion and Future Work}
\label{sec:conclusion}

\ourdataset (StackExchange - Personalized Expert Finding) is an extension of a previous work \cite{se-pqa}, which presents a large real-world dataset for personalized cQA.
The data inherits a rich set of user-level features modeling the  interactions  among the members of the online communities.

Our study provided a detailed description of the data creation and training process. Furthermore, we  illustrated the methodologies adopted, explicitly focusing on IR  techniques. We discussed how the similarity scores computed can be aggregated and combined to target the EF task adopted. 
For the retrieval, we adopted a two-stage  architecture, where the second stage utilizes  for re-ranking an optimized combination of the scores generated by BM25, DistilBERT/MiniLM$_{\text{SBERT}}$, and TAG models.

The preliminary experiments  conducted proved the effectiveness of personalization on this dataset, surpassing methods that rely on pre-trained and fine-tuned large language models by a statistically significant margin.
We expect other researchers to develop more complex strategies to improve results on the \ourdataset resource. We leave such research as future work for us and the IR community working on personalized IR. 

\noindent \textbf{Acknowledgements}. Funding for this research has been provided by: PNRR - M4C2 - Investimento 1.3, Partenariato Esteso PE00000013 - ``FAIR - Future Artificial Intelligence Research'' - Spoke 1 ''Human-centered AI'' funded by the European Union (EU) under the NextGeneration EU programme;  the EU’s Horizon Europe research and innovation programme EFRA (Grant Agreement Number 101093026).  Views and opinions expressed are however those of the author(s) only and do not necessarily reflect those of the EU or European Commission-EU. Neither the EU nor the granting authority can be held responsible for them.
\bibliographystyle{ACM-Reference-Format}
\bibliography{biblio}

\end{document}